\documentclass[
prl,
twocolumn,
preprintnumbers,
superscriptaddress,
longbibliography,
]{revtex4-1}\usepackage{graphicx,amssymb,amsbsy,amsfonts,amssymb,amsmath}
\usepackage{url}
\usepackage[dvipsnames]{xcolor}
\usepackage[hyperindex,breaklinks,pdfusetitle]{hyperref}
\hypersetup{
	colorlinks = true,
	allcolors = Blue,
	linkbordercolor = {white},
}
\usepackage{multirow}
\usepackage{stackrel}
\usepackage{tikz}
\usetikzlibrary{calc}

\newcommand{\be}{\begin{equation}}
\newcommand{\ee}{\end{equation}}

\newcommand{\bea}{\begin{eqnarray}}
\newcommand{\eea}{\end{eqnarray}}

\newcommand{\eps}{\epsilon}

\def\eqn#1{eq.~(\ref{#1})}

\def\nn{\nonumber}

\def\spa#1.#2{\left\langle#1\,#2\right\rangle}
\def\spb#1.#2{\left[#1\,#2\right]}
\def\spash#1.#2{\spa{\smash{#1}}.{\smash{#2}}}
\def\spbsh#1.#2{\spb{\smash{#1}}.{\smash{#2}}}
\def\sand#1.#2.#3{%
\left\langle\smash{#1}{\vphantom1}^{-}\right|{#2}%
\left|\smash{#3}{\vphantom1}^{-}\right\rangle}
\def\sandpp#1.#2.#3{%
\left\langle\smash{#1}{\vphantom1}^{+}\right|{#2}%
\left|\smash{#3}{\vphantom1}^{+}\right\rangle}
\def\sandpm#1.#2.#3{%
\left\langle\smash{#1}{\vphantom1}^{+}\right|{#2}%
\left|\smash{#3}{\vphantom1}^{-}\right\rangle}
\def\sandmp#1.#2.#3{%
\left\langle\smash{#1}{\vphantom1}^{-}\right|{#2}%
\left|\smash{#3}{\vphantom1}^{+}\right\rangle}

\def\ibp{IBP}

\newbox\charbox
\newbox\slabox
\def\s#1{{      
        \setbox\charbox=\hbox{$#1$}
        \setbox\slabox=\hbox{$/$}
        \dimen\charbox=\ht\slabox
        \advance\dimen\charbox by -\dp\slabox
        \advance\dimen\charbox by -\ht\charbox
        \advance\dimen\charbox by \dp\charbox
        \divide\dimen\charbox by 2
        \raise-\dimen\charbox\hbox to \wd\charbox{\hss/\hss}
        \llap{$#1$} }}

\begin{document}

\title{
	The Two-Loop Four-Graviton Scattering Amplitudes
}
\preprint{CP3-20-11, FR-PHENO-2020-002, IPhT-20/003, TTP20-003, MPP-2020-17}

\author{S.~Abreu}
\affiliation{Center for Cosmology, Particle Physics and
	Phenomenology (CP3), Universit\'{e} Catholique de Louvain, 1348
Louvain-La-Neuve, Belgium}
\author{F.~Febres Cordero}
\affiliation{Physics Department, Florida State University
Tallahassee, FL 32306, U.S.A.}
\author{H.~Ita}
\affiliation{Physikalisches Institut,
	Albert-Ludwigs-Universit\"at Freiburg,
D--79104 Freiburg, Germany}
\author{M.~Jaquier}
\affiliation{Institute for Theoretical Particle Physics, KIT, Karlsruhe, Germany.}
\author{B.~Page}
\affiliation{Institut de Physique Th\'eorique, CEA, CNRS, Universit\'e Paris-Saclay, F-91191 Gif-sur-Yvette cedex, France}
\author{M.~S.~Ruf}
\affiliation{Physikalisches Institut,
	Albert-Ludwigs-Universit\"at Freiburg,
D--79104 Freiburg, Germany}
\author{V.~Sotnikov}
\affiliation{Max Planck Insitute for Physics (Werner Heisenberg Institute),
D--80805 Munich, Germany}
\newcommand{\MR}[1]{#1}
\let\Re\relax
\let\Im\relax
\let\imath\relax

\newcommand{\Reals}{\mathbb{R}}
\newcommand{\Complex}{\mathbb{C}}
\newcommand{\imath}{\mathrm{i}}
\newcommand{\loopf}{F_\epsilon}
\begin{abstract}
We present the analytic form of the two-loop four-graviton
scattering amplitudes in Einstein gravity.
To remove ultraviolet divergences we include 		
counterterms quadratic and cubic in the Riemann curvature tensor.
The two-loop numerical unitarity approach is used
to deal with the challenging momentum dependence of the
interactions.
We exploit the algebraic properties of the integrand of the amplitude
in order to reduce it to a minimal basis of Feynman integrals. 
Analytic expressions are obtained from numerical evaluations of the amplitude.
Finally, we show that four-graviton scattering observables depend on 
fewer couplings than naively expected. 

\end{abstract}

\maketitle

Scattering amplitudes are ubiquitous in high-energy physics:
they connect physical observables and
the quantum field theories describing the different 
forces of Nature. By understanding the structure of amplitudes,
we can learn about properties of these theories and their
physical implications. Unlike other field theories,
such as Yang-Mills', Einstein's theory of 
general relativity
cannot be
consistently quantized in its minimal form. 
Indeed, 
it was shown over 30 years ago \cite{Hooft1974,Goroff1985,Goroff1986,Ven1992} that 
quantum effects render scattering amplitudes ill defined in the ultraviolet (UV).
Since then, our understanding of the UV properties has been refined by the study  of 
scattering amplitudes in this regime,
both in Einstein gravity~\cite{Bern2015,Bern2017,Dunbar:2017nfy}
and in supersymmetric extensions of it such as 
maximal supergravity~\cite{Bern2009,Bern2018}. 
New results for amplitudes
have also been obtained, but mostly in
supersymmetric theories~\cite{Green:1982sw,Dunbar:1994bn,Bern1998,Naculich:2008ew,
Boucher-Veronneau2011,Abreu:2019rpt,Chicherin:2019xeg,Henn:2019rgj}.
Computations in Einstein gravity are famously involved, and
while the one-loop four-graviton amplitudes
have been known for decades \cite{Dunbar:1994bn}, 
the two-loop amplitudes remained 
unknown till now. In this letter, we present them for the first time.

Following the detection of gravitational waves, 
interest in quantum gravity amplitudes has surged as a means to
predict the classical gravitational dynamics of large massive
objects in the post-Minkowskian approximation, most
notably that of black-hole binaries
\cite{Damour2016,Damour2018,Cheung2018,Kosower2019,
Antonelli:2019ytb,Bern:2019nnu,Bern:2019crd}. 
Already some time ago, the two-loop scattering amplitudes in string theory 
were understood to yield the classical scattering angle of massless particles
\cite{Amati:1990xe}, but the validity of this observation was recently 
questioned \cite{Damour:2019lcq}. 
Our results give new insights on the theoretical properties of Einstein's theory 
of gravity and associated physical phenomena.
In fact,
the amplitudes presented here were already used~\cite{BernEikonal} 
to confirm the scattering angle of massless particles 
in Einstein gravity~\cite{Amati:1990xe}.

Our calculation is performed with techniques developed for the
computation of amplitudes in the Standard Model of particle physics. They have
already been successfully applied to computations of planar
scattering amplitudes in QCD, both
numerically~\cite{Abreu:2017xsl,Abreu:2017hqn,Abreu:2018jgq} and analytically
\cite{Abreu:2018zmy,Abreu:2019odu}, and are well suited to address the
challenges of a quantum gravity calculation.
We use a variant of the unitarity method
\cite{Bern:1994zx,Bern:1994cg,Britto:2004nc} suitable for numerical
computations, the two-loop numerical unitarity 
approach \cite{Ita:2015tya,Abreu:2017xsl,Abreu:2017idw}, which 
replaces Feynman-diagram input with numerical
evaluations of on-shell tree amplitudes. It bypasses the explicit
construction of the integrand of the amplitude, and directly reduces it to a minimal
basis of Feynman integrals with
unitarity-compatible integration-by-parts relations
\cite{Gluza:2010ws,Schabinger:2011dz}. 
Analytic expressions, provided in a set of ancillary files,
are reconstructed from exact numerical 
evaluations of the amplitudes.

{\flushleft \bf Four-Graviton Scattering Amplitudes.}
We consider four-graviton scattering in Einstein gravity.
The theory is not renormalizable \cite{Hooft1974,Goroff1985,Goroff1986,Ven1992},
and we work in the effective field theory proposed in
ref.~\cite{Donoghue1994}.
The Lagrangian ${\cal L}$  is
\begin{eqnarray}
		\label{eq:fullLag}
		{\cal L}=
		{\cal L}_{\text{EH}}+
		{\cal L}_{\text{GB}}+
		{\cal L}_{\text{R}^3}\, + \ldots,
\end{eqnarray}
where we suppress terms not relevant for our two-loop calculation such as higher-order
operators and those proportional to the equations of 
motion~\cite{Hooft1974,Goroff1985,Goroff1986}.
It is given in terms of the Einstein-Hilbert (EH) Lagrangian ${\cal L}_{\text{EH}}$,
complemented by the Gauss-Bonnet (GB) and the $\text{R}^3$ 
counterterms  \cite{Gibbons:1978ac,Hawking:1979ig,Goroff1985,Goroff1986}, 
denoted ${\cal L}_{\text{GB}}$ and
${\cal L}_{\text{R}^3}$ respectively, whose role is to
cancel the UV divergences inherent to ${\cal L}_{\text{EH}}$.
The different contributions to $\mathcal L$ are
\begin{align}\begin{split}\label{eq:3Lag}
	\!{\cal L}_{\text{EH}}&\!=\!-\frac{2}{\kappa^2} \sqrt{|g|} R\,,\\
		\!{\cal L}_{\text{GB}}&\!=\!\frac{{\cal C}_{\textrm{GB}}}{(4\pi)^{2}} 
		\sqrt{|g|} (R^2\!-\!4R_{\mu\nu}R^{\mu\nu}\!\!+\!\!R_{\mu\nu\rho\sigma}
		R^{\mu\nu\rho\sigma}),\\
		\!{\cal L}_{\text{R}^3}&\!=\!\frac{{\cal C}_{\text{R}^3}}{(4\pi)^{4}}
	\left(\frac{\kappa}{2}\right)^2 \!\!\sqrt{|g|} 
	R_{\alpha\beta}^{\phantom{\alpha\beta}\mu\nu}
	R_{\mu\nu}^{\phantom{\mu\nu}\rho\sigma}
	R_{\rho\sigma}^{\phantom{\rho\sigma}\alpha\beta}\,,
\end{split}\end{align}
where $g=\det(g_{\mu\nu})$ with $g_{\mu\nu}$ 
the metric tensor, $R$ the Ricci scalar, $R_{\mu\nu}$ the
Ricci tensor and $R_{\mu\nu\rho\sigma}$ the Riemann curvature tensor.
We work in the 't Hooft-Veltman (HV) scheme of dimensional regularization, 
with $D=4-2\epsilon$.
So that each contribution has the same dimensions, we introduce the 
dimensionful quantity 
$\mu$ which includes conventional
factors in dimensional regularization,
${\mu}^2=(4\pi)^{-1} e^{\gamma_\mathrm{E}}\mu_0^2$.
The coupling $\kappa$ is related to
Newton's constant $G_\mathrm{N}$,  
$\kappa\mu_0^{-\epsilon}=\sqrt{32\pi G_\mathrm{N}}$. The
divergent parts of the bare couplings
\begin{align}\begin{split}\label{eq:Coeffs}
	{\cal C}_{\textrm{GB}}=&\left(\frac{53}{90}\frac{1}{\epsilon}
	+c_{\textrm{GB}}(\mu)\right){\mu}^{-2\epsilon},\,\\
	{\cal C}_{{\text{R}^3}}=&\left(\frac{209}{1440}\frac{1}{\epsilon}
	+c_{\text{R}^3}(\mu)\right){\mu}^{-4\epsilon}\,,
\end{split}\end{align}
have been determined previously~\cite{Goroff1985,Goroff1986,Bern2015}. 
The renormalized couplings
$c_{\textrm{GB}}(\mu)$ and $c_{\text{R}^3}(\mu)$ will be discussed
at the end of this letter.

We compute graviton scattering on a flat background $\eta_{\mu\nu}$, parametrized 
by the linear split $g_{\mu\nu}=\eta_{\mu\nu}+\kappa h_{\mu\nu}$, 
where  $h_{\mu\nu}$ is the graviton 
field \footnote{We work in the mostly-minus metric signature, 
$\eta_{\mu\nu}=\text{diag}(1,-1,-1,-1)$.}. 
Perturbation theory is defined as an expansion in powers of $\kappa$.
The main results of this letter are
the helicity amplitudes for four-graviton scattering 
$M_{\vec h}(s,t;\eps)$ at order $\kappa^6$, 
as a function of $s=(p_1+p_2)^2$ and $t=(p_2+p_3)^2$, with outgoing momenta $p_i$. 
We will often suppress dependence on Mandelstam variables.
The helicity assignments are specified by
$\vec h=\{h_1,h_2,h_3,h_4\}$.
It is sufficient to compute amplitudes with $\vec h=\{\pm,+,+,+\}$
and $\{-,-,+,+\}$ since all others are related by symmetry.
We define the perturbative expansion of the helicity amplitudes through
\begin{eqnarray}\label{eq:pertExpansion}
{M}_{\vec h}  = \left(\frac{\kappa}{2}\right)^2 {\cal N}_{\vec h} \, \sum_{j\geq0}
\left(\frac{\bar\kappa}{2}\right)^{2j}
 {\cal M}_{\vec h}^{(j)}\,,
\end{eqnarray}
with $\bar{\kappa}=\kappa\mu^{-\epsilon}/(4\pi)$ and
helicity-dependent phases ${\cal N}_{\vec h}$ given in footnote~\footnote{
We define the phase factors
${\cal N}_{++++}=\imath\left[\!\frac{[12]}{\langle 12\rangle}\!\frac{[34]}{\langle 34\rangle}\!\right]^2$,
${\cal N}_{-+++}=\imath\left[\!\frac{\langle 14\rangle}{[14]}\!\frac{[24]}{\langle 24\rangle}\!\frac{[34]}
{\langle 34\rangle}\!\right]^2$ and 
${\cal N}_{--++}=\imath\left[\!\frac{\langle 12\rangle}{[12]}\!\frac{[34]}{\langle 34\rangle}\!\right]^2$,
where we used spinor helicity notation (see e.g.~\cite{Maitre:2007jq}).
}.
That is, we normalize ${M}_{\vec h}$ so that the coefficients 
${\cal M}_{\vec h}^{(j)}$ are Lorentz invariant.
The index~$j$
in eq.~\eqref{eq:pertExpansion} is in one-to-one correspondence with the loop-order
of the contributing diagrams for ${\cal L}_{\text{EH}}$. This correspondence
breaks down for ${\cal L}_{\text{GB}}$ and ${\cal L}_{\text{R}^3}$
as can be seen by the power of the coupling in the three-point
vertices of each term in eq.~\eqref{eq:3Lag}:
a three-point vertex is $\mathcal{O}(\kappa)$ in ${\cal L}_{\text{EH}}$,
$\mathcal{O}(\kappa^3)$ in ${\cal L}_{\text{GB}}$ and 
$\mathcal{O}(\kappa^5)$ in~${\cal L}_{\text{R}^3}$.
This implies that ${\cal M}_{\vec h}^{(2)}$ has tree, one-loop, and two-loop contributions, 
depending on which vertex appears. Schematically,
\begin{align}\begin{split}\nn
&{\cal M}_{\vec h}^{(0)}\sim\raisebox{-2.5mm}{\includegraphics[scale=0.17]{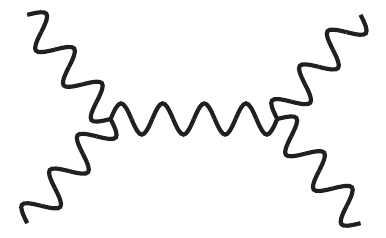}}+\ldots\,,\quad
{\cal M}_{\vec h}^{(1)}\sim\raisebox{-2.5mm}{\includegraphics[scale=0.15]{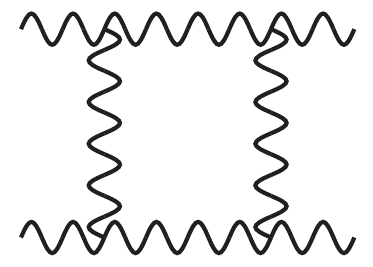}}+\ldots\,,\\
&{\cal M}_{\vec h}^{(2)}\sim
\raisebox{-2.5mm}{\includegraphics[scale=0.15]{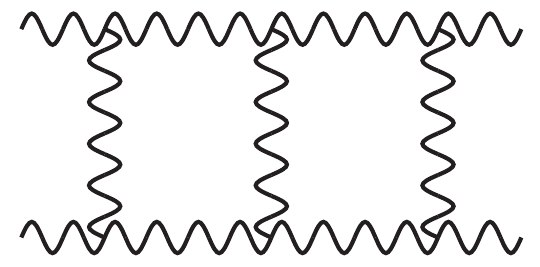}}+
\raisebox{-3mm}{\includegraphics[scale=0.17]{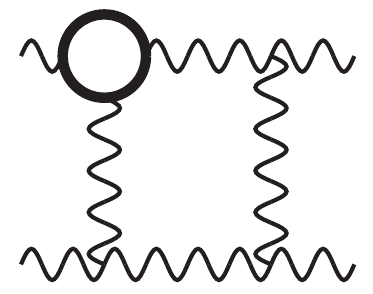}}+
\raisebox{-3mm}{\includegraphics[scale=0.2]{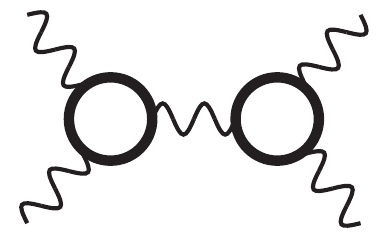}}+
\raisebox{-3mm}{\includegraphics[scale=0.2]{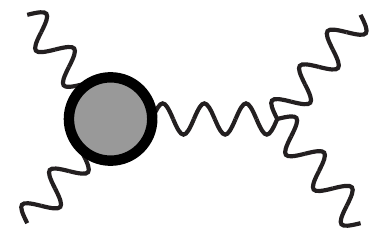}}+\ldots,
\end{split}\end{align}
where we include sample diagrams for each contribution.
White blobs denote ${\cal L}_{\text{GB}}$ vertices, grey blobs denote 
${\cal L}_{\text{R}^3}$ vertices and ${\cal L}_{\text{EH}}$ vertices have no
decoration. The first non-vanishing contributions from 
${\cal L}_{\text{GB}}$ and ${\cal L}_{\text{R}^3}$ appear at ${\cal O}(\kappa^{6})$.

The amplitudes ${\cal M}_{\vec h}^{(j)}$ computed from $\mathcal{L}$ in eq.~\eqref{eq:fullLag}
are UV finite, but there remain infrared (IR) 
singularities~\cite{Weinberg1965,Naculich:2011ry,Naculich:2013xa, Akhoury:2011kq}. 
It is known \cite{Bern:1998ug} that there are no 
collinear singularities, and the soft singularities exponentiate.
We define
\begin{equation} \label{eq:soft}
  {\cal S} = 
    \sum_{i<j}^4 \frac{\mu_0^{2\epsilon}}{\epsilon^2}
    (-(p_i+p_j)^2)^{1-\epsilon}
    \,,
\end{equation}
and construct 
finite functions ${\cal F}^{(j)}_{\vec h}(\epsilon)$ through
\begin{equation}
  \begin{split}
M_{\vec h} =  
\left(\frac{\kappa}{2}\right)^2 {\cal N}_{\vec h} \, &\mathrm{exp}\left[\left(\frac{\bar\kappa}{2}\right)^2 {\cal S} \right] 
\sum_{j\geq0}
\left(\frac{\bar\kappa}{2}\right)^{2j}{\cal F}_{\vec h}^{(j)}(\epsilon)\,.
  \label{eq:rem}
  \end{split}
\end{equation}
Comparing eqs.~\eqref{eq:pertExpansion} and \eqref{eq:rem}, we
can write the ${\cal F}_{\vec h}^{(j)}(\epsilon)$ in terms of the
${\cal M}_{\vec h}^{(j)}$ and $\cal S$.
The two-loop remainder is 
\begin{equation}
	{\cal R}_{\vec h}^{(2)}
	={\cal F}_{\vec h}^{(2)}(0)
	=\lim_{\epsilon\to0}\!\left(\!
	{\cal M}_{\vec h}^{(2)}-{\cal S}{\cal M}_{\vec h}^{(1)}+\frac{{\cal S}^2}{2}{\cal M}_{\vec h}^{(0)}
	\!\right).
  \label{eq:remfinal}
\end{equation}
This object captures the new four-dimensional information at two-loops.

{\flushleft\bf Computation.}
The main obstacles in computing the amplitudes 
${\cal M}_{\vec h}^{(2)}$ are rooted in the 
involved Feynman rules derived from $\mathcal{L}$ in eq.~\eqref{eq:fullLag}. 
Vertices have many terms with high powers of the momenta, making it
hard to construct the integrand
of the amplitude. Furthermore, despite the simple kinematics of the process,
the reduction to a set of master
integrals is challenging because the integrand has high powers of the loop
momentum. 

\begin{figure}[ht] \begin{tikzpicture}[scale=1.2]
\node at (4.6,3.3){\includegraphics[scale=0.13]{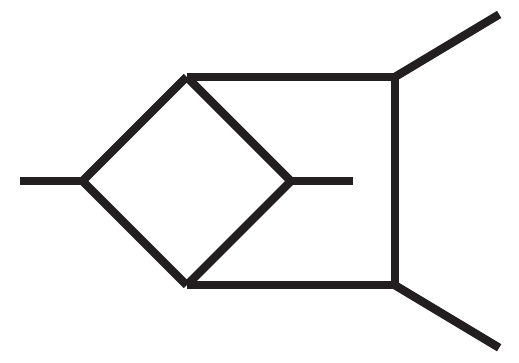}}; 
\node at (3.4,3.3){\includegraphics[scale=0.13]{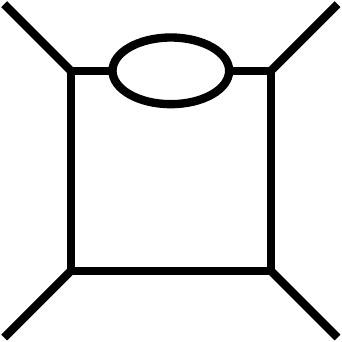}}; 
\node at (2.2,3.3){\includegraphics[scale=0.13]{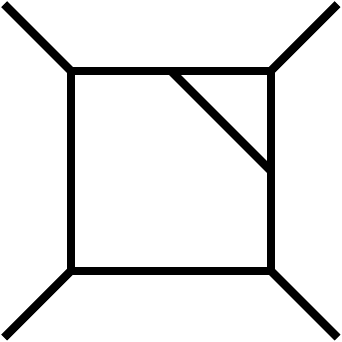}}; 
\node at (1,3.3){\includegraphics[scale=0.13]{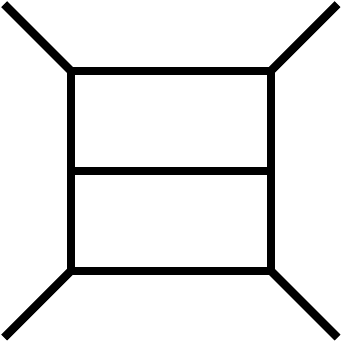}};
\node at (0,2.4){\includegraphics[scale=0.13]{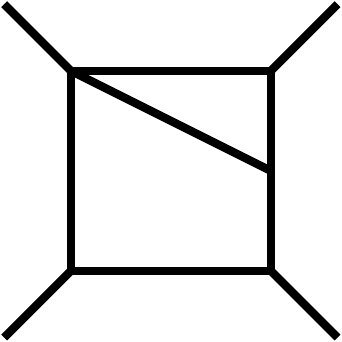}}; 
\node at (0.65,2.4){\includegraphics[scale=0.13]{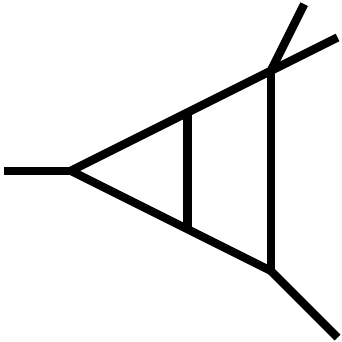}}; 
\node at (1.3,2.4){\includegraphics[scale=0.13]{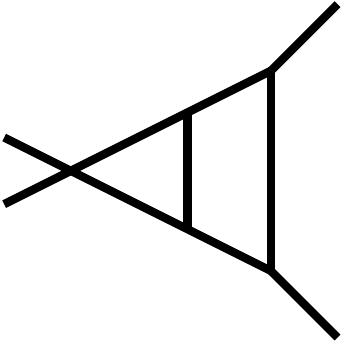}};
\node at (2,2.4){\includegraphics[scale=0.13]{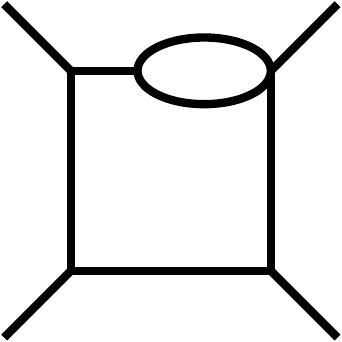}};
\node at (2.7,2.4){\includegraphics[scale=0.13]{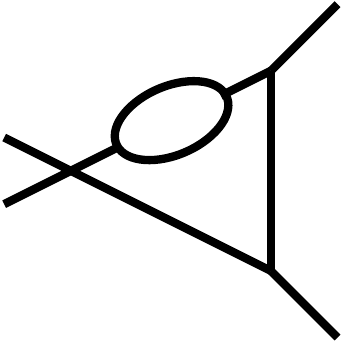}};
\node at (3.4,2.4){\includegraphics[scale=0.13]{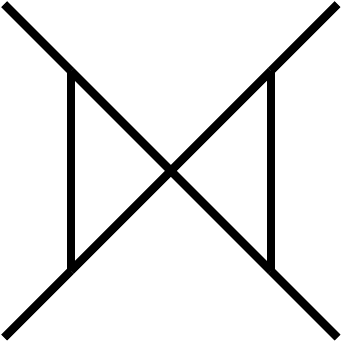}};
\node at (4.05,2.4){\includegraphics[scale=0.13]{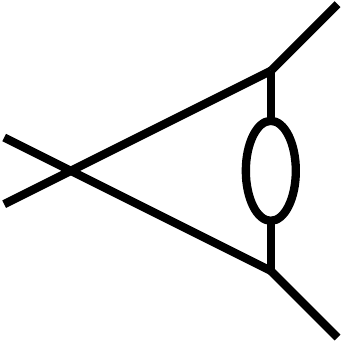}};
\node at (4.8,2.45){\includegraphics[scale=0.13]{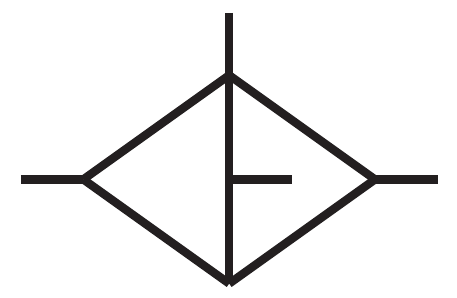}};
\node at (5.65,2.394){\includegraphics[scale=0.13]{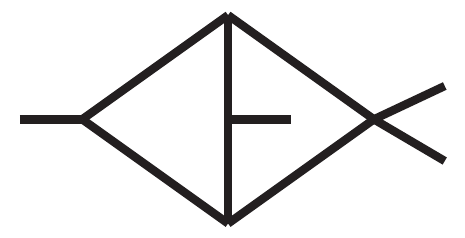}};
\node at (0,1.4){\includegraphics[scale=0.13]{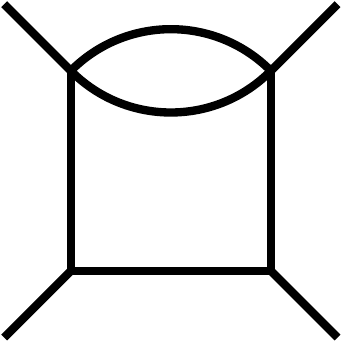}}; 
\node at (.7,1.4){\includegraphics[scale=0.13]{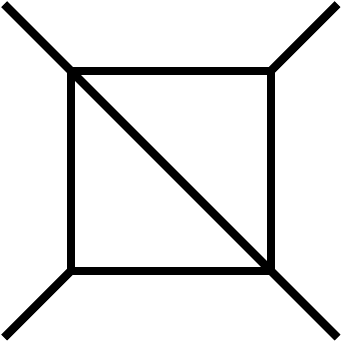}};
\node at (1.4,1.4){\includegraphics[scale=0.13]{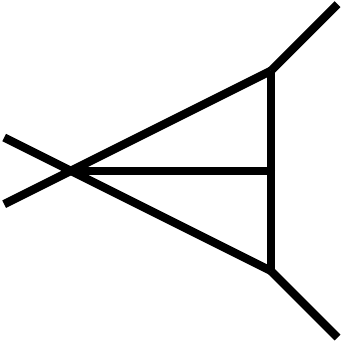}};
\node at (2.1,1.4){\includegraphics[scale=0.13]{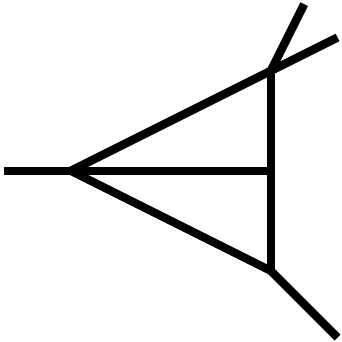}};
\node at (2.8,1.4){\includegraphics[scale=0.13]{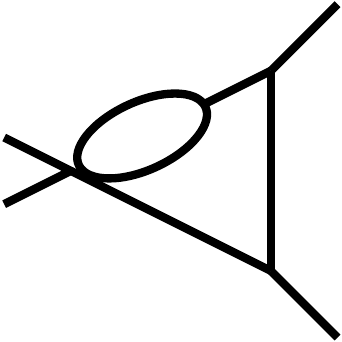}};
\node at (3.5,1.4){\includegraphics[scale=0.14,angle=180]{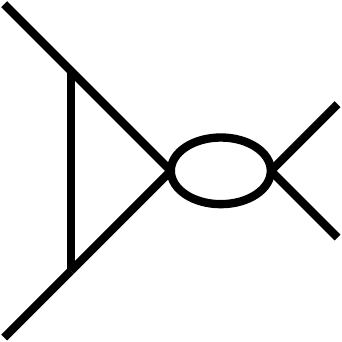}};
\node at (4.2,1.4){\includegraphics[scale=0.13]{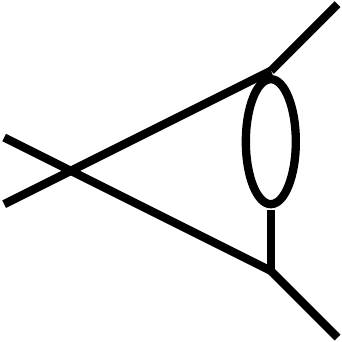}};
\node at (4.9,1.4){\includegraphics[scale=0.14]{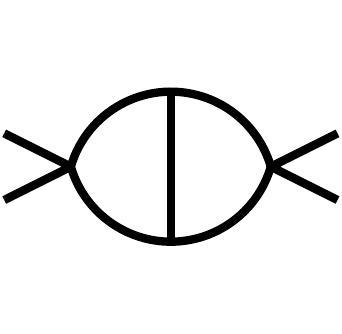}};
\node at (5.65,1.4){\includegraphics[scale=0.14]{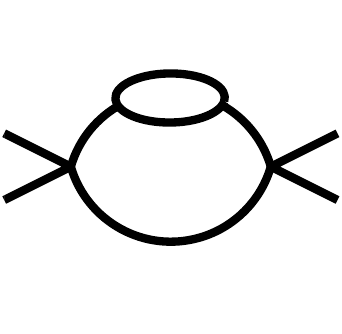}};
    \node at 
(4.6,.6){\includegraphics[scale=0.13]{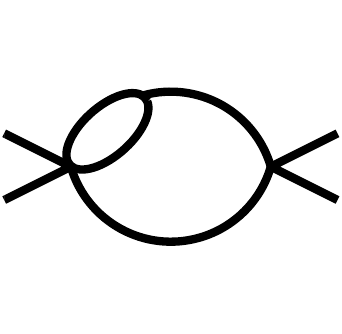}};
    \node at
(3.4,.6){\includegraphics[scale=0.13]{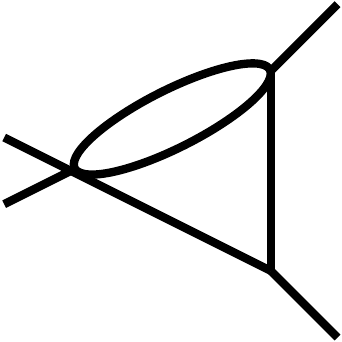}};
\node at 
(2.2,.6){\includegraphics[scale=0.13]{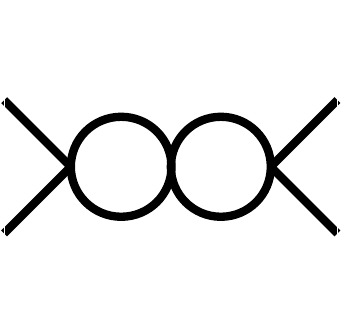}};
\node at
(1,.6){\includegraphics[scale=0.13]{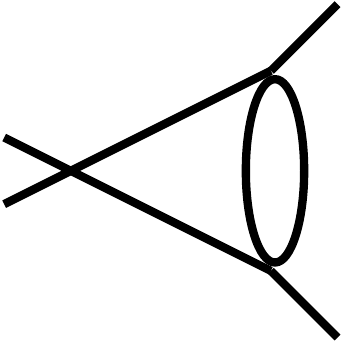}}; 
    \node at (2.8,-0.1){\includegraphics[scale=0.15]{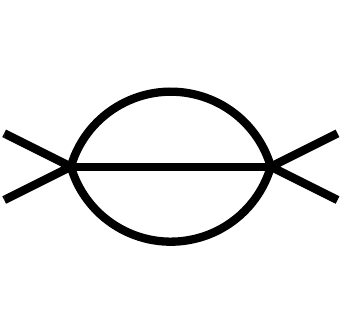}};
\end{tikzpicture} \caption{Topologically inequivalent propagator structures
for 2-to-2 scattering, including non-planar contributions.}
\label{fig:DiagramsHierarchy}
\end{figure}

The framework of two-loop numerical unitarity 
\cite{Ita:2015tya,Abreu:2017xsl,Abreu:2017idw} 
is particularly well suited to address these challenges.
The starting point is the following parametrization of the integrand of an 
amplitude \cite{Ita:2016oar}, denoted $\mathcal{M}^{(k)}(\ell_l)$,
\begin{equation}\label{eq:AL}
    \mathcal{M}^{(k)}(\ell_l)=\sum_{\Gamma\in\Delta}
    \sum_{i\in M_\Gamma\cup S_\Gamma} c_{\Gamma,i}
    \frac{m_{\Gamma,i}(\ell_l)}{\prod_{j\in
    P_\Gamma}\rho_j}\, ,
\end{equation}
with $M_\Gamma$ a set of master integrands, $S_\Gamma$ a set of surface terms,
$P_\Gamma$ the set of propagators $\rho_j$ associated with each propagator 
structure $\Gamma$, and $\ell_l$ the set of loop momenta. 
The set $\Delta$ of relevant propagator structures is characterized
in fig.~\ref{fig:DiagramsHierarchy}.
The undetermined
coefficients $c_{\Gamma,i}$ in the decomposition \eqref{eq:AL} are constrained from the factorization properties
of the integrand in loop-momenta configurations $\ell^\Gamma_l$ where
the propagators in $P_\Gamma$ vanish:
\begin{equation}
\sum_{\rm states}\prod_{k\in T_\Gamma}
{\cal M}^{\rm tree}_k(\ell_l^\Gamma) =\!\!\!\!
\sum_{\substack{\Gamma' \ge \Gamma\,,\\ 
i\,\in\,M_{\Gamma'}\cup S_{\Gamma'}}} \!\!\!\!
\frac{ c_{\Gamma',i}\,m_{\Gamma',i}(\ell_l^\Gamma)}
{\prod_{j\in (P_{\Gamma'}\setminus P_\Gamma) } \rho_j(\ell_l^\Gamma)}\,,
\label{eq:CE}
\end{equation}
with $T_\Gamma$ the tree amplitudes corresponding to the vertices in $\Gamma$.
The sum over states runs over $D_\mathrm{s}$-dimensional graviton helicity states,
and the sum over $\Gamma'$ runs over propagator structures such that 
$P_{\Gamma}\subseteq P_{\Gamma'}$. 
The system of eqs.~\eqref{eq:CE}
is constructed numerically. Assuming we have built the decomposition \eqref{eq:AL}
and can evaluate the product of trees in \eqref{eq:CE},
this reduces the calculation of the amplitudes 
at a phase-space point to solving the linear system of eqs.~\eqref{eq:CE}.
Indeed, once all $c_{\Gamma,i}$ have been determined, we directly obtain the decomposition 
of the amplitude in terms of  master integrals,
\begin{equation}\label{eq:AI}
    \mathcal{M}^{(j)}=\sum_{\Gamma\in\Delta}
    \sum_{i\in M_\Gamma} c_{\Gamma,i} I_{\Gamma,i}\,,
\end{equation}
where the integrals $I_{\Gamma,i}$ correspond to the master integrands in $M_\Gamma$.
In the following, we discuss the construction of decomposition \eqref{eq:AL}
and the computation of tree-amplitudes for eqs.~\eqref{eq:CE}.

We first focus on the evaluation of tree amplitudes. We use a
fast numerical algorithm provided
by Berends-Giele recursion~\cite{Berends:1987me}. In the pure Einstein-Hilbert 
theory, $\mathcal{L}_{\textrm{EH}}$, we use the reformulation in terms of 
cubic interactions proposed in ref.~\cite{Cheung2017}.
For counterterm contributions,
vertices are computed using the 
program~{\sc xAct}~\cite{xAct,Brizuela:2008ra,Nutma:2013zea}.
Our Berends-Giele recursion allows for EH, GB and $\text{R}^3$ tree amplitudes. 
We use integer values for the state
counting parameter $D_\mathrm{s}$ that are large enough to recover the full
momentum dependence, i.e.~$D_\mathrm{s}\geq6$.

Next, we discuss the construction of the decomposition~\eqref{eq:AL}. 
It depends on the power-counting properties of the theory and 
the kinematics of the process. 
First, we build the full set of propagator structures $\Delta$, which contains both
planar and non-planar contributions, see fig.~\ref{fig:DiagramsHierarchy}.
For each $\Gamma\in \Delta$ we then construct the function space
$M_\Gamma\cup S_\Gamma$. The elements of the space, $m_{\Gamma,i}(\ell_l)$, are
polynomials in the components of the loop momenta $\ell_l$. The linear span of
the space is controlled by the  theory-specific maximal polynomial degree. 
In Einstein gravity one naively expects that the
polynomial degree required is twice that of Yang-Mills. The next step is the
construction of the surface terms in $S_\Gamma$, which integrate to zero. 
A subset of these can
be built from tensor reduction techniques \cite{Abreu:2017xsl}. The rest are
constructed from integration-by-parts (IBP) relations
\begin{equation}
0 = \int \prod_{l=1,2} \mathrm{d}^D\ell_l \frac{\partial}{\partial \ell_{i}^\nu} \left[ 
    \frac{ u_{i}^\nu }{\prod_{k\in P_\Gamma} \rho_k}\right],  \label{eq:ibp} 
\end{equation} 
provided that 
\begin{equation} 
\label{eq:GKK} u_{i}^{\nu} \frac{\partial}{\partial \ell_{i}^\nu} \rho_{j}= 
f_j \rho_j,  
\end{equation}
so that no new higher propagator powers are generated in the 
procedure~\cite{Gluza:2010ws,Schabinger:2011dz}.
The $f_j$ are polynomials in loop-momenta components, and
no summation over the index $j$ is implied. 
Solutions $u_i^\nu$ to \eqn{eq:GKK} are power-counting independent and referred to as \ibp{}-generating vectors.
For each $\Gamma$, once a set of vectors is found, surface terms are constructed as follows.
Consider a polynomial $t_r(\ell_l)$ in the loop-momenta components 
and a solution $u^{\nu}_{i,s}$ to eq.~\eqref{eq:GKK}. We then insert 
$t_r(\ell_l)u^{\nu}_{i,s}$ in eq.~\eqref{eq:ibp} to obtain the surface term
\begin{equation} 
\label{eq:surfaceTerm} 
m_{\Gamma,(r,s)} \!= 
 u_{i,s}^{\nu} \frac{\partial t_r(\ell_l)}{\partial \ell_i^\nu} 
+ t_r(\ell_l) \left(\frac{\partial u_{i,s}^{\nu}}{\partial \ell_i^\nu} 
-\sum_{k\in P_\Gamma} f^s_{k}\right),
\end{equation}
where $D$-dependence may arise from the divergence term.
We complete the \ibp-generating vectors obtained in ref.~\cite{Abreu:2017xsl} for 
planar topologies with the ones for non-planar topologies.
To obtain surface terms with the suitable power-counting, we
must use a sufficient set of polynomials $t_r(\ell)$. 
Each vector $u^\nu_i$ 
appears in many surface terms, offering the
opportunity for caching in the numerical approach. 
The set of master integrands $M_\Gamma$ in eq.~\eqref{eq:AL} is the
complement of $S_\Gamma$ in the integrand function space.

We are now ready to
construct the system of eqs.~\eqref{eq:CE}. For each numerical
phase-space point, choice of $\epsilon=(4-D)/2$, and value of $D_\mathrm{s}$, we can solve
for the coefficients $c_{\Gamma,i}$, yielding the decomposition \eqref{eq:AI} in terms of
master integrals. 
To expand the result in
$\epsilon$, we first reconstruct the dependence of the coefficients on this
parameter and~$D_\mathrm{s}$. They are rational in $\epsilon$, 
and so we compute a sufficient number of samples to apply
Thiele's formula~\cite{abramowitz1964handbook}. The coefficients of $\epsilon$ in the
numerator of this rational function depend on $D_\mathrm{s}$. In pure gravity, they
are quartic polynomials in $D_\mathrm{s}$. 
GB counterterm amplitudes have rational  $D_\mathrm{s}$ dependence,
with numerators that are cubic in $D_\mathrm{s}$ and denominators that are simply 
$D_\mathrm{s}-2$~\footnote{This can be understood to be the combination of two terms:
a quadratic polynomial in $D_\mathrm{s}$ and a quadratic polynomial in $D_\mathrm{s}$ divided
by $D_\mathrm{s}-2$. The pole at $D_\mathrm{s}=2$ is introduced by the EH propagator 
in axial gauge~\cite{Capper:1981rc}. Its contributions is projected out in pure gravity 
but not in the presence of a GB vertex. The quadratic numerators are the most generic $D_\mathrm{s}$ 
dependence we can have in a one-loop gravity amplitude.}. 
The $\text{R}^3$ counterterm amplitudes are $D_\mathrm{s}$ independent. We determine the 
$D_\mathrm{s}$ dependence from enough numerical samples.

Through this procedure, we obtain master integral coefficients as rational functions in 
$\epsilon$, with analytic $D_\mathrm{s}$ dependence at numerical values of $s$ and $t$.
We set $D_\mathrm{s}=4-2\eps$, as prescribed by the HV scheme, 
insert the expressions for the master integrals \cite{Smirnov:1999gc,Tausk:1999vh,Smirnov:1999wz,Anastasiou:2000kp},
and expand the result in $\epsilon$.
With modern mathematical tools \cite{Duhr:2019tlz} we can express the amplitudes 
in a basis $B$ of classical polylogarithms, whose elements are denoted $h_i\in B$. 
Using one-loop amplitudes we computed within the same framework, we obtain
the remainders in \eqref{eq:remfinal} at the chosen phase-space point
as a linear combination of the $h_i$:
\begin{equation}\label{eq:remainderDec}
	\mathcal{R}^{(2)}_{\vec h}(s,t)=\sum_{h_i \in B}d_i(s,t)\,h_i(s,t)\,.
\end{equation}

Finally, we can reconstruct the full analytic
result from a sufficient number of numerical samples. As noted e.g.~in
refs.~\cite{Abreu:2018zmy,Badger:2018enw,Abreu:2019odu,Badger:2019djh}, it is
more efficient to reconstruct the
coefficients $d_i(s,t)$ of eq.~\eqref{eq:remainderDec}.
The coefficients $d_i$ are rational functions of $x=t/s$, and the $s$
dependence can be reconstructed from dimensional analysis.
Therefore, we can use the univariate Thiele formula to reconstruct
the rational functions $d_i$. This process requires around 20 numerical
samples for each helicity.
Numerical stability issues are sidestepped by employing
finite-field arithmetic~\cite{vonManteuffel:2014ixa,Peraro:2016wsq}. 
Combining the results from evaluations over two different finite
fields with cardinality of order~$2^{31}$,
we lift the results to the field of rational numbers using
the Chinese remainder theorem and rational
reconstruction techniques \cite{wang1981p}.

{\flushleft\bf Results.}
We have computed the four-graviton amplitude for the three independent helicity
configurations $\vec h=\{-,-,+,+\}$, $\{-,+,+,+\}$ and $\{+,+,+,+\}$, up to
order ${\cal O}(\kappa^6)$ in the effective field theory of eq.~\eqref{eq:fullLag}.
The amplitudes are obtained by computing the remainders of 
eq.~(\ref{eq:remfinal}) and then reinstating the IR singularities. 
By taking into account the contributions from the
GB (up to one-loop) and the tree-level $\text{R}^3$ counterterms,
we also obtain the two-loop amplitudes in the EH
theory. We note that the evaluation of the remainders requires one-loop
amplitudes through ${\cal O}(\epsilon)$.
All scattering amplitudes, in the HV scheme, are provided in ancillary files.

We performed several checks on our results. 
First we verified that all the poles in our amplitudes, which
are by construction of IR origin, are accounted for by
the universal structure~\eqref{eq:soft}.
The absence of UV poles directly confirms the UV divergences computed in ref.~\cite{Goroff1985,Goroff1986}.
Second, some parts of the different ingredients we require to compute the 
${\cal O}(\kappa^6)$ amplitudes have been obtained previously,
giving completely independent checks.
One-loop amplitudes in
Einstein gravity  were computed in ref.~\cite{Dunbar:1994bn} through ${\cal O}(\epsilon^0)$. 
We confirm the $\{\pm,+,+,+\}$ results,
the $\{-,-,+,+\}$ amplitude up to a sign
\footnote{We match \cite{Dunbar:1994bn} after a sign flip of
the $A^{\mathcal{N}=1}$ sub-amplitude in the $\{-,-,+,+\}$ helicity configuration.} and
agree with an independent computation~\cite{BernUnpub}.
The counterterm amplitudes were partially known. 
We reproduce the divergent pieces of
the counterterm amplitudes for $\{+,+,+,+\}$ given in 
ref.~\cite{Bern2015}.
The GB tree-level and one-loop amplitudes match an independent computation
of the $\{+,+,+,+\}$ and $\{-,-,+,+\}$ helicities \cite{BernUnpub}.
Regarding $\text{R}^3$, we reproduce known results for the tree-level amplitudes with a single 
$\text{R}^3$ insertion \cite{Bern2015,Dunbar:2017qxb}.
Third, we could check some of the ${\cal O}(\kappa^6)$ amplitudes:
the $\{+,+,+,+\}$ amplitude matches the results of ref.~\cite{BernUnpub}
and is consistent with ref.~\cite{Dunbar:2017qxb},
and our results for the $\{-,-,+,+\}$ amplitude match the
behaviour established in~\cite{Bartels:2012ra}.
Finally, our amplitudes behave consistently with factorization in the
limits where the Mandelstam invariants $s$, $t$ or $u=-s-t$ vanish.

While the results are too large to print in this letter, we can quote the result for the $\{-,-,+,+\}$ remainder
in the $s$-channel Regge limit. Defined by $s\gg -t>0$, this limit is directly
relevant for linking scattering amplitudes to classical dynamics \cite{BernEikonal}. 
With our choice of IR subtraction, we 
find~\footnote{We note that two numerical coefficients in eq.~\eqref{eqn:remainder}
have changed with respect to the first version of the letter. In the first
version, we had wrongly set the dimensional regulator in the one-loop
four-graviton amplitudes which amounted to using $D_s=4+4\eps$ instead of
$D_s=4-2\eps$ in this part. This led to an inconsistent shift in the
$\mathcal{O}(\eps)$ terms of the one-loop four-graviton amplitudes 
and an induced shift of the finite remainders.}
\begin{align}\label{eqn:remainder}\begin{split}
	&\mathcal{R}^{(2)}_{\{-,-,+,+\}}=
	s^3\left\{2\frac{s}{t}\pi^2\left(\frac{\imath\pi}{2}-L\right)^2
	-3 \pi ^2L^2\right.\\
	&+\frac{107}{10} \pi ^2 L+\frac{14191}{1350} \pi ^2-
	\frac{158 }{45}\pi^4-\frac{13049}{2160}\\
	&+\imath \pi  \left[-\frac{14 }{3}L^3+\frac{87}{10}L^2-\left(8 \pi ^2-\frac{9437}{225}\right)L\right.\\
	&\quad\left.\left.-20 \zeta_3+\frac{2621 }{210}\pi ^2-\frac{25567}{750}\right]+\mathcal{O}(-t/s)\right\}\,,
\end{split}\end{align}
where we introduced  $L=\log(-s/t)$. This expression is independent
of the scale $\mu$ introduced in eq.~\eqref{eq:3Lag}, consistent with the following
discussion.

Finally, we consider the remainders' dependence on the couplings in 
eq.~\eqref{eq:Coeffs}. 
${\cal R}^{(2)}_{\{-,-,+,+\}}$ is independent of both
$c_{\textrm{GB}}(\mu)$ and $c_{{\text{R}^3}}(\mu)$, while  
${\cal R}^{(2)}_{\{\pm,+,+,+\}}$ depend on the unique combination
\begin{align} \label{eq:couplingComb}
	c(\mu)=c_{{\text{R}^3}}(\mu)-\frac{1}{2}\,c_{{\text{GB}}}(\mu)\,.
\end{align}
This observation is tightly connected with the dependence of the 
remainders on the scale $\mu$ introduced in eq.~\eqref{eq:3Lag}. Indeed
we find that
\begin{align} \label{eqn:RG}
 	\mu\frac{\partial}{\partial \mu}{\cal R}_{\vec{h}}^{(2)}={}&
	\left[\frac{1}{120}
	+ \mu\frac{\partial}{\partial \mu} c(\mu)\right]
	\frac{\mathcal{M}^{{\rm tree},\mathrm{R}^3}_{\vec{h}}}{{\cal C}_{\text{R}^3} } \,,
\end{align}
where $\mathcal{M}^{{\rm tree},\mathrm{R}^3}_{\vec{h}}$ is the tree amplitude 
with a single $\mathrm{R}^3$ insertion, which vanishes for 
$\vec h=\{-,-,+,+\}$. 
This extends the 
scale dependence proposed for $\vec h=\{+,+,+,+\}$ in refs.~\cite{Bern2015,Bern2017,Dunbar:2017nfy}
to all helicities. 
The scale dependence in eq.~\eqref{eqn:RG} takes a much simpler numerical 
form than the divergent parts of the couplings in eq.~\eqref{eq:Coeffs}.
Requiring that the remainders are independent
of $\mu$ allows to determine the $\mu$-dependence of the coupling $c(\mu)$
\footnote{The solution $c(\mu)=1/120\log (\lambda/\mu)$ requires to introduce the 
scale $\lambda$. Together with $\kappa$,
they are the two physical parameters of the effective field theory \eqref{eq:fullLag}.}. 
This is sufficient for the remainders to be well defined,
and it is a weaker condition than requiring that 
${\cal C}_{\textrm{GB}}$ and ${\cal C}_{{\text{R}^3}}$ be $\mu$-independent.

The fact that remainders display a reduced dependence on the couplings
in eq.~\eqref{eq:Coeffs} is interesting for two reasons. First,
the same is not true regarding how the couplings contribute to the
cancellation of the UV poles. This yields \emph{two} independent equations,
allowing to uniquely fix the divergent part of the couplings 
${\cal C}_{\text{R}^3}$ and ${\cal C}_{\textrm{GB}}$.
Second, this implies that  physical observables related to four-graviton scattering at two-loops
depend on fewer parameters than those appearing in the effective field theory.
It is likely that this degeneracy is a consequence of
the evanescence of the GB counterterm, which would then imply that our observation
should extend to two-loop amplitudes of higher multiplicities. This is consistent
with the results of ref.~\cite{Dunbar:2017qxb}, which can be shown to imply that the two-loop five-point
all-plus amplitude depends on the same combination $c(\mu)$ of couplings.

{\flushleft\bf Conclusions.}
In this letter, we presented the ${\cal O}(\kappa^6)$ four-graviton amplitudes in Einstein
gravity, including contributions from counterterms. The computation of graviton
amplitudes is notoriously difficult but our results show that modern
field-theory methods, notably the numerical unitarity approach, are able to
tackle these challenges.
Our results give new insights into the analytic structure of
the theory, contributing~\cite{BernEikonal} to the ongoing effort to bridge 
multi-loop scattering amplitudes and classical gravitational dynamics.
We find that the $\{-,-,+,+\}$ remainder only depends on the coupling $\kappa$, 
while the $\{\pm,+,+,+\}$ amplitudes depend on a single additional coupling. 
This implies that observables constructed from these remainders only depend
on two out of the three couplings appearing in the effective field theory.

Multiple future directions are worth pursuing. Given the mild dependence of our
approach on the number of scales, a clear next step is to consider
amplitudes including massive particles. Another natural extension is towards
higher loop corrections. 
Both will be of direct relevance for exploring 
the classical gravitational dynamics of large massive objects.
The analytic results we present also provide insights into the
analytic properties of the amplitudes, stimulating the development of
more efficient techniques to tackle calculations at higher
loop orders and multiplicities.

\section{Acknowledgments}

We thank Z.~Bern, C.~Duhr, H.~Johansson, C.~Steinwachs and M.~Zeng for many helpful discussions.
We thank
Z.~Bern, C.~Cheung, H.-H.~Chi, S.~Davies, L.~Dixon and J.~Nohle for sharing 
unpublished results for the counterterm amplitudes and the all-plus remainder \cite{BernUnpub}.
The work of S.A.~is supported by the Fonds de la
Recherche Scientifique--FNRS, Belgium.
S.A.~wishes to thank CERN's theory department for its hospitality.
The work of F.F.C.\ is supported by the 
U.S. Department of Energy under grant DE-SC0010102.
The work of V.S.\ is supported by the European Research Council (ERC) 
under the European Union's Horizon 2020 research and innovation programme,
\textit{Novel structures in scattering amplitudes} (grant agreement No.\ 725110).
H.I. thanks the Pauli Center of ETH Z\"urich and the University of Z\"urich for hospitality.          
The work of B.P.~is supported by the French Agence Nationale
pour la Recherche, under grant ANR--17--CE31--0001--01.
M.S.R.'s work is funded by the German Research Foundation (DFG) within the Research Training Group GRK 2044.
M.S.R.~wishes to thank the ETH Z\"urich for its hospitality. 
This research was
supported by the Munich Institute for Astro- and Particle Physics (MIAPP) of
the DFG cluster of excellence ``Origin and Structure of the Universe''.
This work used computational and storage services associated with the Hoffman2 
Shared Cluster provided by UCLA Institute for Digital Research and Education's 
Research Technology Group.
The authors acknowledge support by the state of
Baden-W\"urttemberg through bwHPC.
\bibliography{main.bib}
\end{document}